\documentclass[12pt]{article}
\usepackage{epsf}
\def\beq{\begin{equation}}
\def\eeq{\end{equation}}

\def\pp{\psi(2S)}
\def\jp{J/\psi}
\def\ypp{\Upsilon(2S)}
\def\yp{\Upsilon}
\def\yppp{\Upsilon(3S)}

\begin{document}
\begin{titlepage}
\begin{center}
{\Large \bf William I. Fine Theoretical Physics Institute \\
University of Minnesota \\}
\end{center}
\vspace{0.2in}
\begin{flushright}
TPI-MINN-06/22-T \\
UMN-TH-2508-06 \\
June 2006 \\
\end{flushright}
\vspace{0.3in}
\begin{center}
{\Large \bf Two-pion transitions in quarkonium revisited
\\}
\vspace{0.2in}
{\bf M.B. Voloshin  \\ }
William I. Fine Theoretical Physics Institute, University of
Minnesota,\\ Minneapolis, MN 55455 \\
and \\
Institute of Theoretical and Experimental Physics, Moscow, 117218
\\[0.2in]
\end{center}

\begin{abstract}
Two-pion transitions in charmonium and bottomonium are considered with the
leading relativistic effects taken into account. The contribution of the chromo-
magnetic interaction of the charmed quarks to the amplitude of the decay
$\psi(2S) \to \pi \pi J/\psi$ is estimated from the available data. This
contribution is enhanced by a factor of three in the decay $\eta_c(2S) \to \pi
\pi \eta_c$ and should produce a noticeable modification of the rate and the
spectrum in the latter decay. It is argued that the peculiar observed
spectrum in the decay $\Upsilon(3S) \to \pi \pi \Upsilon$ arises due to a
dynamical suppression of the leading nonrelativistic quarkonium amplitude and
thus enhanced prominence of the relativistic terms. Also discussed
are the effects of the final state interaction between the pions.
\end{abstract}

\end{titlepage}

\section{Introduction}
Hadronic transitions between states of the available heavy quarkonia, charmonium
and bottomonium, present a very interesting case study in dynamics of both the
heavy quarks and the light mesons emitted in the transitions. In particular, the
observed properties of such processes with the emission of two pions are
understood within the chiral low-energy dynamics of pions, starting with the
earliest observation~\cite{spear} of the decay $\pp \to \pi^+ \pi^- \jp$ and the
theoretical analyses of the data~\cite{mv0,bc}. Furthermore, the QCD picture of
such decays is that the heavy quarkonium transition generates soft gluonic field
which then produces the light mesons. The heavy quarkonium can be considered as
a compact and nonrelativistic object in its interaction with a soft gluonic
field, which justifies the use of the multipole expansion in QCD in analysing
these processes~\cite{gottfried}. In this approach the amplitude of the process
factorizes into the heavy quarkonium part, i.e. the transition between the
levels as a source of the field, and the light meson part, which describes the
creation of the mesons by the field operators. The former part depends on the
dynamics of the quarkonium, while the latter one can be understood in some
detail by combining~\cite{vz} the methods of chiral dynamics and the general
low-energy relations in QCD, in particular those for the anomalies in the trace
of the energy-momentum tensor and in the singlet axial current. Each of these
two factors in the amplitudes of the discussed hadronic transitions can be used,
in certain extent, independently. For instance, the decays $\pp \to \pi^+ \pi^-
\jp$ and $\pp \to \eta \jp$ (as well as $\ypp \to \pi^+ \pi^- \yp$ and  $\ypp
\to \eta \yp$) have essentially the same heavy quarkonium part, so that the
ratio of the decay rates is fully determined by the low-energy meson
amplitudes~\cite{vz,mv2}. On the other hand the charmonium amplitude for the
double interaction with the chromo-electric field (the chromo-polarizability),
which dominates the quarkonium factor in the decay $\pp \to \pi^+ \pi^- \jp$ can
be used~\cite{sv} for considering the low-energy limit of the conversion of
$\pp$ into $\jp$ on a nucleon: $\pp + N \to \jp +N$, where the soft gluonic
field is that inside the nucleon. The same charmonium transition amplitude can
be used as a benchmark for the diagonal chromo-polarizability of $\jp$, which
determines the low-energy elastic scattering of $\jp$ on a nucleon~\cite{kv,sv}.
For these reasons a more detailed understanding of both the heavy quarkonium
amplitudes and of those determining the gluon conversion into light mesons is
desirable.

Presently a quite detailed experimental data are available on the dominant
hadronic transitions in charmonium~\cite{bes}: $\pp \to \pi^+ \pi^- \jp$, and in
bottomonium~\cite{cleo1,cleo2}:  $\ypp \to \pi^+ \pi^- \yp$ and  $\yppp \to
\pi^+ \pi^- \Upsilon(1S,2S)$. Numerous other, less visible, transitions have
been observed, e.g. $\Upsilon(1 D) \to \pi^+ \pi^- \yp$~\cite{cleo3}. Thus it is
not unreasonable to expect, given the capabilities of the modern $e^+e^-$
experiments at the charm and bottom thresholds, that a significantly improved
data will become available. Such data may enable a study of finer effects in the
amplitudes of the hadronic transitions beyond the leading ones thus far
considered in the literature.

The purpose of the present paper is to analyze sub-leading relativistic effects
in the amplitudes of the two-pion transitions in charmonium and bottomonium, and
possible ways of studying such effects from the data. For the transitions
between the $^3S_1$ states of quarkonium, dominated by the second order in the
leading $E1$ term of the multipole expansion, these corrections arise from the
second order in the $M1$ interaction. It will be argued that in the charmonium
transition $\pp \to \pi^+ \pi^- \jp$ the correction adds coherently to the
leading term as an $O(10\%)$ term. Being of a rather moderate value in this
particular decay, the correction is enhanced by a factor of 3 in the transition
$\eta_c(2S) \to \pi^+ \pi^- \eta_c$ with a potentially very noticeable effect in
the spectrum of the dipion invariant mass. Furthermore, this correction modifies
the estimate of the charmonium chromo-polarizability with implications for
charmonium scattering on nucleons. It will be further argued that, although
generally the relativistic effects in bottomonium are quite small, in the
particular case of the decay $\yppp \to \pi^+ \pi^- \yp$ it appears that the
leading nonrelativistic term it strongly suppressed dynamically, and including
the relativistic effects of the $M1$ interaction and of the $^3D_1 - ^3S_1$
mixing can possibly resolve the long standing puzzle of the peculiar observed
spectrum of the dipion invariant masses in this decay, which spectrum by far
does not conform with the predictions from the leading chromo-electric
interaction.

Also some consideration will be given to the significance of the final state
interaction (FSI) between the pions in the two-pion transitions. The FSI effects
have been discussed ever since the earliest analyses~\cite{bc}, and are
generally believed to be  moderate in the kinematical region of the decays $\pp
\to \pi^+ \pi^- \jp$ and $\ypp \to \pi^+ \pi^- \yp$. However once the amplitudes
of the transitions are studied in finer detail, one will have to better quantify
these effects. Although at present the FSI effects cannot be fully analyzed,
their behavior at low invariant mass of the two pions can be estimated within
the chiral expansion. For this purpose we give here a systematic derivation of
the amplitudes for two pion production by gluonic operators in the leading
chiral limit, including the terms with the pion mass, and consider the first
sub-leading term due to iteration of the chiral amplitudes. At the invariant
mass of the two pions near the threshold these corrections are quite small -
only few percent. Their continuation to higher values of the invariant mass can
be studied experimentally in a rather straightforward way. Such study, in
particular, would quantify the impact of the FSI effects on the estimates of the
quarkonium amplitudes from the observed decay rates. It will be also argued that
even with the already available data such impact can be estimated as quite
moderate, as generally expected.~\footnote{In particular, the present data
certainly exclude the recently claimed~\cite{guo} modification due to FSI of the
charmonium amplitude by a factor of about 3.}

The material in the present paper is organized as follows. In Section 2 the
chromo-electric and chromo-magnetic dipole interactions responsible for the
two-pion transitions are discussed and the parameters determining the relevant
quarkonium amplitudes to order $v^2/c^2$ are introduced. The Section 3 contains
a compilative derivation in the leading chiral order of the amplitudes for
creation of the pion pair by gluonic operators, and in Section 4 the discussion
of the two previous sections is combined in the expressions for the two-pion
transition amplitudes with the leading relativistic terms included. The latter
expressions are confronted with the available date on the transitions $\pp \to
\pi \pi \jp$ and $\ypp \to \pi \pi \yp$ in Section 5 and the significance of the
chromo-magnetic interaction in transitions between charmonium resonances is
evaluated. In Section 6 the FSI effects are considered using iteration of the
chiral amplitudes and it is also argued on phenomenological grounds that these
effects are only of a moderate magnitude in the observed two-pion transitions.
The Section 7 addresses the long-standing puzzle of the unusual spectrum of the
dipion invariant masses observed in the decay $\yppp \to \pi \pi \yp$ and it is
suggested that this spectrum can be explained if the formally leading
nonrelativistic quarkonium amplitude is dynamically suppressed in this
transition, so that the relativistic terms provide a significant contribution.
Finally, Section 8  contains a summary of the discussion in the present paper.

\section{Multipole expansion for the two-pion transitions and relativistic
terms}

Considering quarkonium as a compact nonrelativistic object one can apply the
multipole expansion for the quarkonium interaction with soft gluonic
field~\cite{gottfried,mv1}. The leading term in this expansion is the
chromo-electric dipole interaction with the chromo-electric field ${\vec E}^a$
described by the Hamiltonian
\beq
H_{E1}=-{1 \over 2} \xi^a \, {\vec r} \cdot {\vec E}^a (0)~,
\label{e1}
\eeq
where $\xi^a=t_1^a-t_2^a$ is the difference of the color generators acting on
the quark and antiquark (e.g. $t_1^a = \lambda^a/2$ with $\lambda^a$ being the
Gell-Mann matrices), and ${\vec r}$ is the vector for relative position of the
quark and the antiquark. The convention used throughout this paper is that the
QCD coupling $g$ is included in normalization of the gluon field operators, so
that e.g. the gluon field Lagrangian is written as $- (F_{\mu \nu}^a)^2/(4g^2)$.

The amplitude of the two-pion transition between heavy quarkonium states $\psi_2
\to \pi^+ \pi^- \psi_1$ (with $\psi_2$ and $\psi_1$ used as a generic notation
for the initial and the final quarkonium states) can thus be written
as~\footnote{For definiteness the processes with emission of a pair of charged
pions, $\pi^+\pi^-$ are considered here, since the emission of neutral pions
$\pi^0\pi^0$ is trivially related by the isospin symmetry.}
\beq
A(\psi_2 \to \pi^+ \pi^- \psi_1) = {1 \over 2} \langle \pi^+ \pi^- | E^a_i E^a_j
| 0 \rangle \, \alpha^{(12)}_{i j}~,
\label{e1ampg}
\eeq
where $\alpha^{(12)}_{i j}$ can be termed, in complete analogy with the atomic
properties in electric field, as the transitional chromo-polarizability of the
quarkonium. In other words, the $\psi_2 \to \psi_1$ transition in the
chromo-electric field is described by the effective Hamiltonian
\beq
H_{eff}=-{1 \over 2}\, \alpha^{(12)}_{i j} \, E^a_i E^a_j~,
\label{heff}
\eeq
with the chromo-polarizability given by
\beq
\alpha^{(12)}_{i j}={1 \over 16} \, \langle \psi_1 | \xi^a \, r_i \, {\cal G} \,
r_j \, \xi^a | \psi_2 \rangle~,
\label{aij}
\eeq
where ${\cal G}$ is the Green's function of the heavy quark pair in a color
octet state. The latter function is not well understood presently, so that an
{\it ab initio} calculation of the chromo-polarizability would be at least
highly model dependent.

Also a simple remark is in order that the discussed decays in quarkonia are
governed by transitional polarizabilities, i.e. those linking different states
of quarkonium. The diagonal chromo-polarizability of quarkonium states, in
particular of the charmonium resonances $\pp$ and $\jp$ can also be
measured~\cite{mv3} and is relevant to the problem of scattering of these
resonances on nuclei~\cite{sv}.

Generally $\alpha_{i j}$ is a symmetric tensor. Clearly, for transitions between
pure $S$ wave quarkonium states this tensor is necessarily proportional to
$\delta_{ij}$:  $\alpha_{i j} = \alpha_0 \, \delta_{ij}$. However for the
$^3S_1$ states a $^3D_1 - ^3S_1$ mixing generally takes place due to the
relativistic effects in the order $v^2/c^2$ so that a traceless $D$ wave part of
the chromo-polarizability is also present and is proportional to spin-2
combination of the polarizations, ${\vec \psi}_2$ and ${\vec \psi}_1$  of the
initial and the final states:
\beq
\alpha_{i j} = \alpha_0 \, \delta_{ij} \, \, ({\vec \psi}_1 \cdot {\vec \psi}_2)
+ \alpha_2\, \left [ \psi_{1i} \psi_{2j} + \psi_{1j} \psi_{2i} - (2/3) \,
\delta_{ij} \, ({\vec \psi}_1 \cdot {\vec \psi}_2) \right ]~.
\label{asd}
\eeq
Numerically, the $^3D_1 - ^3S_1$ mixing in charmonium appears to be comparable
with the characteristic value of the relativistic parameter for this system,
$v^2/c^2 \approx 0.2$. In particular, Rosner~\cite{rosner} using the $e^+e^-$
decay width of $\psi(3770)$ and considering this resonance as dominantly a
$^3D_1$ state, estimates the angle $\theta$ of the $\psi(3770) - \pp$ mixing,
i.e. the $^3D_1 - ^3S_1$ mixing for $\pp$, as $\theta \approx 0.2$. It can be
noticed however that the mixing does not contribute to the spinless part
$\alpha_0$ of the chromo-polarizability and that in the spin-averaged transition
rate there is no interference between the $S$ and $D$ wave parts of
$\alpha_{ij}$. Thus the effect of the $^3D_1 - ^3S_1$ mixing in the rate is of
order $v^4/c^4$ and is thus generally very small.

It should be also mentioned that the effects of the recoil of the final
quarkonium state in the decay also arise in the order $v^2/c^2$ and contribute
terms quadratic in the momentum of the pion pair ${\vec q}$ to the spin-diagonal
part of the tensor polarizability $\alpha_{ij}$. However such terms would also
be of higher order in the chiral expansion in the pion momenta, which is used
here. Therefore the recoil terms are suppressed by a product of two small
parameters and can be neglected within the approximations adopted in this
paper.~\footnote{The recoil effects in the quarkonium amplitude were considered
in Ref.~\cite{bdm} and estimated to be quite small.}

Besides the discussed $^3D_1 - ^3S_1$ mixing, the only other effect in the order
$v^2/c^2$ in the amplitudes of two-pion transitions between quarkonium $S$ wave
states arises through the second order in the M1 interaction with the
chromo-magnetic field ${\vec B}^a$ described by the Hamiltonian
\beq
H_{M1}= - {1 \over 2
\, M}\, \xi^a \, ({\vec \Delta} \cdot {\vec B}^a)~,
\label{m1}
\eeq
where ${\vec \Delta}={\vec s}_1-{\vec s}_2$ is the difference of
the spin operators acting on the quark and the antiquark, and $M$ is the heavy
quark mass. The contribution of this term to the amplitude of two-pion
transition between the $S$ states of quarkonium can be written as
\begin{eqnarray}
\label{am1}
&&A_M(\psi_2 \to \pi^+ \pi^- \psi_1) =
\\ \nonumber
&& {1 \over 2} \langle \pi^+\pi^-| B_i^a B_j^a |0 \rangle \, \alpha_M^{(12)} \,
\left [  \delta_{ij} \, ({\vec \psi}_1 \cdot {\vec \psi}_2) + {3 \over 2} \,
(\psi_{1i} \psi_{2j} + \psi_{1j} \psi_{2i} - (2/3) \, \delta_{ij} \, {\vec
\psi}_1 \cdot {\vec \psi}_2 ) \right ]~,
\end{eqnarray}
for the transitions between $^3S_1$ states, and
\beq
A_M(\eta_2 \to \pi^+ \pi^- \eta_1) =  {3 \over 2} \langle \pi^+\pi^-| B_i^a
B_i^a |0 \rangle \, \alpha_M^{(12)}
\label{am0}
\eeq
for the transitions between $^1S_0$ states. (The notation $\eta_2$ and $\eta_1$
is used here for the states of quarkonium in order to emphasize that this
relation is specific for the $^1S_0$ states.) The amplitude $\alpha_M^{(12)}$ is
a chromo-magnetic analog of the chromo-electric term $\alpha_{ij}$ and is
formally given by the formula
\beq
\alpha_M^{(12)}={1 \over 48 M^2} \, \langle \phi_1 | \xi^a \, {\cal G} \xi^a |
\phi_2 \rangle~,
\label{amg}
\eeq
where $\phi_2$ and $\phi_1$ are the coordinate parts of the $S$-wave wave
functions of the initial and the final states. It is taken into account here
that the amplitudes generated by the M1 interaction are already suppressed in
comparison with those described by Eq.(\ref{e1ampg}) by the factor $v^2/c^2$.
Therefore the wave functions of the quarkonium states can be taken in the form
where the spin and coordinate parts are factorized.

The spin-2 part of the amplitude in Eq.(\ref{am1}) contributes to the decay rate
only in the order $v^4/c^4$, similarly to the effect of the $^3D_1-^3S_1$
mixing. However, the spin-0 part in this amplitude as well as that in
Eq.(\ref{am0}) (the only one present there) does interfere with the leading
nonrelativistic amplitude in Eq.(\ref{e1ampg}) proportional to $\alpha_0$ from
Eq.(\ref{asd}). Therefore this part provides the first relativistic correction,
proportional to $v^2/c^2$, to the transition rate. One can also notice that the
effect for the $^1S_0$ states is three times bigger than for the $^3S_1$ states.

\section{Two-pion creation by gluonic operator. Chiral limit.}

As one can see from the previous section, a crucial role in the discussed
approach to calculating the two-pion transition amplitudes is played by the
amplitudes for production of two pions by operators quadratic in components of
the gluon field strength tensor. Therefore
in this section we consider the general amplitude of such type: $\langle
\pi^+(p_1) \pi^-(p_2)| F_{\mu \nu}^a F_{\lambda \sigma}^a|0 \rangle$,
describing the creation of two pions by the local gluonic operator $F_{\mu \nu}
F_{\lambda \sigma}$. In the leading chiral limit the momenta $p_1$ and $p_2$ of
the pions as well as the pion mass $m$ are to be considered as small parameters,
and the expression for the amplitude, quadratic in these parameters, can be
written in the following general form
\beq
- \langle \pi^+(p_1) \pi^-(p_2)| F_{\mu \nu}^a F_{\lambda \sigma}^a|0 \rangle =
\left [ X\, (p_1 \cdot p_2) + Y \,(p_1^2+p_2^2- m^2) \right ] \, (g_{\mu
\lambda} g_{\nu \sigma} - g_{\mu \sigma} g_{\nu \lambda})+ Z \, t_{\mu \nu
\lambda \sigma}~,
\label{genf}
\eeq
where the structure
\begin{eqnarray}
t_{\mu \nu \lambda \sigma}&=&(p_{1 \mu} p_{2\lambda}+p_{1 \lambda} p_{2 \mu}) \,
g_{\nu \sigma} +(p_{1 \nu} p_{2\sigma}+p_{1 \sigma} p_{2 \nu}) \, g_{\mu
\lambda} \nonumber \\ &-& (p_{1 \mu} p_{2\sigma}+p_{1 \sigma} p_{2 \mu}) \,
g_{\nu \lambda}  - 
(p_{1 \nu} p_{2\lambda}+p_{1 \lambda} p_{2 \nu}) \, g_{\mu \sigma} - (p_1 \cdot
p_2) \,(g_{\mu \lambda} g_{\nu \sigma} - g_{\mu \sigma} g_{\nu \lambda})
\label{gent}
\end{eqnarray}
has zero overall trace: ${t_{\mu \nu}}^{\mu \nu}=0$, and $X$, $Y$, and $Z$ are
yet undetermined coefficients. The form of the amplitude in Eq.(\ref{genf}) is
uniquely determined by the symmetry (with respect to the indices) of the
operator $F_{\mu \nu} F_{\lambda \sigma}$ and by the Adler zero condition, which
requires that the amplitude goes to zero when either one of the two pion momenta
is set to zero and the other one is on the mass shell\footnote{The proper index
symmetry and the Adler zero condition also automatically ensure that the
amplitude is C even, i.e. symmetric under the permutation of the pion momenta:
$p_1 \leftrightarrow p_2$.}.

The coefficients $X$ and $Y$ in Eq.(\ref{genf}) are in fact determined~\cite{vz}
by the anomaly in the trace of the energy-momentum tensor $\theta_{\mu \nu}$ in
QCD. Indeed, in the low-energy limit in QCD, i.e. in QCD with three light
quarks, one finds
\beq
\theta_\mu^\mu=-{b \over 32 \pi^2} \, F^a_{\mu \nu} F^{a \, \mu
\nu}+\sum_{q=u,d,s} m_q (\bar q q)~,
\label{anom}
\eeq
where $b=9$ is the first coefficient in the beta function for QCD with three
quark flavors. The first term in Eq.(\ref{anom}) represents the conformal
anomaly, while the quark mass term arises due to the explicit breaking of the
scale invariance by quark masses. On the other hand, the matrix element of the
energy-momentum tensor $\theta_{\mu \nu}$ over the pions: $\theta_{\mu
\nu}(p_1,p_2) \equiv\langle \pi^+(p_1) \pi^-(p_2)|\theta_{\mu \nu}|0 \rangle$ is
fully determined~\cite{vz,ns,dv} in the quadratic in $p_1, p_2$ and $m$ order by
the conditions of symmetry in $\mu$ and $\nu$, conservation on the mass shell
($(p_1+p_2)^\mu \, \theta_{\mu \nu}(p_1,p_2) =0$ at $p_1^2=p_2^2=m^2$),
normalization  ($\theta_{\mu \nu}(p,-p)=2\, p_\mu p_\nu$ at $p^2=m^2$), and the
Adler zero condition ($\theta_{\mu \nu}(p,0)|_{p^2=m^2}=0$):
\beq
\theta_{\mu \nu}(p_1,p_2)=\left [(p_1 \cdot p_2) +p_1^2 +p_2^2 -m^2 \right ]\,
g_{\mu \nu} -p_{1 \mu} p_{2\nu}- p_{1 \nu} p_{2 \mu}~.
\label{st}
\eeq
The equations (\ref{genf}) and (\ref{st}) allow for the pion momenta to be
off-shell in order to demonstrate the Adler zero. However
in what follows only the amplitudes with pions on the mass shell will be
considered, so that it will be henceforth implied that $p_1^2=p_2^2=m^2$.
In particular one finds for the trace of the expression in Eq.(\ref{st})
\beq
\theta_\mu^\mu(p_1,p_2)=2\,(p_1 \cdot p_2) + 4\,m^2~.
\label{tst}
\eeq
Furthermore, the quark mass term in Eq.(\ref{anom}), corresponding to the
explicit breaking of the chiral symmetry in QCD corresponds to the same symmetry
breaking by the pion mass term in the pion theory. Thus one finds to the
quadratic order in $m^2$:
\beq
\langle \pi^+ \pi^- |\sum_{q=u,d} m_q (\bar q q)| 0 \rangle=m^2~,
\label{mt}
\eeq
while the term with the strange quark makes no contribution to the discussed
amplitude.

Combining the formula in Eq.(\ref{anom}) for $\theta_\mu^\mu$ with the
expressions (\ref{tst}) and (\ref{mt}) one finds the matrix element of the
gluonic operator over the pions in the form\footnote{It can be mentioned that
this relation, taking into account the pion mass, was used in Ref.~\cite{mv},
and was also derived in a particular chiral model in Refs.~\cite{ccgm,ccggm}.}
\beq
- \langle \pi^+(p_1) \pi^-(p_2)| F_{\mu \nu}^a F^{a\, \mu \nu}|0 \rangle =
{32 \pi^2 \over b} \, \left [ 2\,(p_1 \cdot p_2) + 3 \,m^2 \right ]~
\label{gt}
\eeq
which thus allows to determine the coefficients $X$ and $Y$ in Eq.(\ref{genf}):
$X=16 \pi^2/(3 b)$ and $Y=3 X/2= 8 \pi^2/b$.

The coefficient $Z$ of the traceless part in Eq.(\ref{genf}) cannot be found
from the trace relation (\ref{anom}). Novikov and Shifman~\cite{ns} estimated
this coefficient by relating this part to the matrix element of the traceless
(twist-two) energy-momentum tensor of the gluons only: $\theta_{\mu \nu}^G = 4
\pi \alpha_s \, (- F_{\mu \lambda}^a {F_{\nu}^a}^\lambda + {1 \over 4} \, g_{\mu
\nu} \, F_{\lambda \sigma}^a F^{a\, \lambda \sigma} )$,
\beq
Z\, {t_{\mu \lambda \nu}}^\lambda = 4 \pi \alpha_s \,  \langle \pi^+(p_1)
\pi^-(p_2)|\theta_{\mu \nu}^G |0 \rangle~.
\label{nsr}
\eeq
They then assume that the matrix element of the twist-two operator is
proportional to the traceless part of the phenomenological energy momentum
tensor of the pions,
\beq
\langle \pi^+(p_1) \pi^-(p_2)|\theta_{\mu \nu}^G |0 \rangle= \rho_G \, \left [
{1 \over 2} \, (p_1 \cdot p_2) \, g_{\mu \nu} - p_{1 \mu} p_{2\nu}- p_{1 \nu}
p_{2 \mu} \right ]
\label{rhog}
\eeq
with the proportionality coefficient  interpreted, similarly to the deep
inelastic scattering, as ``the fraction of the pion momentum carried by gluons".
They further introduce a related parameter $\kappa = b \alpha_s \rho_G/(6 \pi)$
and conjecture that numerically $\kappa \approx 0.15 - 0.20$.

For the purpose of the present consideration the interpretation of the parameter
$\kappa$ as being related to the pion gluon structure function is not essential
and we treat $\kappa$ here as a phenomenological and measurable, and actually
measured, parameter.  Summarizing the results so far in this section one can
write the expression for the general matrix element (\ref{genf}) for on-shell
pions as
\beq
- \langle \pi^+(p_1) \pi^-(p_2)| F_{\mu \nu}^a F_{\lambda \sigma}^a|0 \rangle =
{8 \pi^2 \over 3 b} \,\left [ (q^2+m^2) \, (g_{\mu \lambda} g_{\nu \sigma} -
g_{\mu \sigma} g_{\nu \lambda})- {9 \over 2} \, \kappa \, t_{\mu \nu \lambda
\sigma} \right ] ~,
\label{kapf}
\eeq
where $q=p_1+p_2$ is the total four-momentum of the dipion.

Few remarks are due regarding effects of higher order in $\alpha_s$. The trace
term in Eq.(\ref{kapf}) receives no renormalization, provided that the
coefficient $b$ is replaced by $\beta(\alpha_s)/\alpha_s^2$ with
$\beta(\alpha_s)=b \, \alpha_s^2 + O(\alpha_s^3)$ being the full beta function
in QCD. This modification however only affects the overall normalization of the
trace part, and can in fact be absorbed into the definition of the heavy
quarkonium amplitudes. On the contrary, the relative coefficient of the
traceless term in Eq.(\ref{kapf}), i.e. the parameter $\kappa$, does depend on
the normalization scale, which scale is appropriate to be chosen as the
characteristic size of the heavy quarkonium~\cite{ns}. However, given other
uncertainties in the analysis of the two-pion transitions, the  logarithmic
variation of $\kappa$ is a small effect. In particular, this effect is likely to
be smaller than the discussed in this paper relativistic effects in the
amplitudes of the two-pion emission.

The matrix element in Eq.(\ref{kapf}) describes the production of the two pions
in two partial waves in their center of mass system: the $S$ wave and the $D$
wave. The two waves can be measured separately, and also any effects of the
final state interaction between pions are different in these two orbital states.
Therefore it is quite instructive for the subsequent discussion to explicitly
separate the $S$ and $D$ waves in the matrix element, i.e. to rewrite the
amplitude (\ref{kapf}) in the form
\beq
- \langle \pi^+(p_1) \pi^-(p_2)| F_{\mu \nu}^a F_{\lambda \sigma}^a|0
\rangle=S_{\mu \nu \lambda \sigma}+ D_{\mu \nu \lambda \sigma}~.
\label{sd}
\eeq

Clearly, the trace term in Eq.(\ref{kapf}) corresponds to a pure $S$ wave, while
the traceless term proportional to $\kappa$ contains both waves. In order to
perform explicit partial wave separation in $t_{\mu \nu \lambda \sigma}$ it is
helpful to introduce~\cite{ns} the four vector $r=p_1-p_2$ describing the
relative momentum of the two pions, which  reduces to a purely spatial vector in
the c.m. system of the pions ($(r \cdot q)=0$). Then the tensor
\beq
\ell_{\mu \nu} = r_\mu r_\nu + {1 \over 3} \, \left ( 1- {4 m^2 \over q^2}
\right ) \, (q^2 \, g_{\mu \nu}- q_\mu q_\nu)
\label{lmn}
\eeq
is also purely spatial in the c.m. frame and corresponds to pure $D$ wave. Using
this tensor one can make the following series of replacements for the terms of
the generic structure $p_{1\alpha} p_{2 \beta}$ in the tensor $t_{\mu \nu
\lambda \sigma}$:
\begin{eqnarray}
p_{1\alpha} p_{2 \beta} & \to & {1 \over 4} \, q_\alpha q_\beta - {1 \over 4} \,
r_\alpha r_\beta = {1 \over 4} \, q_\alpha q_\beta  +{1 \over 12} \, \left ( 1-
{4 m^2 \over q^2} \right ) \, (q^2 \, g_{\alpha \beta}- q_\alpha q_\beta)- {1
\over 4} \, \ell_{\alpha \beta}
\nonumber \\
& \to & {1 \over 6} \, \left ( 1+ {2 m^2 \over q^2} \right ) \, q_\alpha q_\beta
- {1 \over 4} \, \ell_{\alpha \beta}~.
\label{ppsub}
\end{eqnarray}
Here in the first replacement the cross terms between $r$ and $q$ are dropped
since they cancel in $t_{\mu \nu \lambda \sigma}$ due to the C symmetry ($p_1
\leftrightarrow p_2$), while the $g_{\alpha \beta}$ term in the last transition
is dropped, since such structure cancels in the traceless tensor $t$. Using
Eq.(\ref{ppsub}) one readily finds from the formula (\ref{kapf}) the expressions
for the $S$ and $D$ wave amplitudes:
\begin{eqnarray}
\label{sw}
&& S_{\mu \nu \lambda \sigma}={8 \pi^2 \over 3 b} \, \left \{ (q^2+m^2)\,
(g_{\mu \lambda} g_{\nu \sigma} - g_{\mu \sigma} g_{\nu \lambda})\right .
 \\  \nonumber
&& \left . - {3 \over 2} \, \kappa \, \left ( 1+ {2 m^2 \over q^2} \right ) \,
\left [ q_\mu q_\lambda g_{\nu \sigma}+q_\nu q_\sigma g_{\mu \lambda}-q_\nu
q_\lambda g_{\mu \sigma}-
q_\mu q_\sigma g_{\nu \lambda}-{1 \over 2} \, q^2\,(g_{\mu \lambda} g_{\nu
\sigma} - g_{\mu \sigma} g_{\nu \lambda})\right ] \right \}~,
\end{eqnarray}
and
\beq
D_{\mu \nu \lambda \sigma}={8 \pi^2 \over 3 b} \,{9 \kappa \over 4}\, \left (
\ell_{\mu \lambda} g_{\nu \sigma}+\ell_{\nu \sigma} g_{\mu \lambda}-\ell_{\nu
\lambda} g_{\mu \sigma}-
\ell_{\mu \sigma} g_{\nu \lambda} \right )~.
\label{dw}
\eeq

\section{Two-pion transition amplitudes with the relativistic corrections}

Using the formulas in the equations (\ref{e1ampg}), (\ref{asd}) and (\ref{am1})
and the expressions (\ref{sw}) and (\ref{dw}) for the dipion production
amplitudes in the chiral limit, one can readily find the amplitude of the
transition $\psi_2 \to \pi^+ \pi^- \psi_1$ between generic $^3S_1$ states of a
heavy quarkonium. After a straightforward calculation one finds the $S$ wave
decay amplitude
\begin{eqnarray}
\label{spp}
&&S(\psi_2 \to \pi^+ \pi^- \psi_1)= \\ \nonumber
&&-{4 \pi^2 \over b} \, \alpha_0^{(12)} \, \left [ (1-\chi_M) \, (q^2+m^2)-
(1+\chi_M) \, \kappa \, \left (1+{2 m^2 \over q^2} \right)\,  \left ( {(q\cdot
P)^2 \over P^2} - {1 \over 2} \, q^2 \right ) \right ] \,(\psi_1 \cdot \psi_2)~,
\end{eqnarray}
as well as three types of $D$ wave amplitude: one unrelated to the spins of the
quarkonium states
\beq
D_1(\psi_2 \to \pi^+ \pi^- \psi_1)=-{4 \pi^2 \over b} \, \alpha_0^{(12)} \,
(1+\chi_M) \, {3 \kappa \over 2} \, {\ell_{\mu \nu} P^\mu P^\nu \over P^2} \,
(\psi_1 \cdot \psi_2)~,
\label{d1}
\eeq
and two amplitudes with the correlation with the polarization of the initial and
the final resonances
\beq
D_2(\psi_2 \to \pi^+ \pi^- \psi_1)= {4 \pi^2 \over b} \, \alpha_0^{(12)} \,
\left (\chi_2 + {3 \over 2} \, \chi_M \right ) \, {\kappa \over 2} \, \left ( 1
+ {2 m^2 \over q^2} \right)\, q_\mu q_\nu \psi^{\mu \nu}~
\label{d2}
\eeq
and
\beq
D_3(\psi_2 \to \pi^+ \pi^- \psi_1)= {4 \pi^2 \over b} \, \alpha_0^{(12)} \,
\left (\chi_2 + {3 \over 2} \, \chi_M \right ) \, {3 \kappa \over 4} \,
\ell_{\mu \nu} \psi^{\mu \nu}~.
\label{d3}
\eeq
In these formulas the following notation is used: $P$ stands for the 4-momentum
of the initial quarkonium resonance, $\psi_1^\mu$ and $\psi_2^\mu$ are the
polarization 4-vectors for the $^3S_1$ states, and $\psi^{\mu \nu}$ is the
spin-2 structure $\psi^{\mu \nu}=\psi_1^\mu \psi_2^\nu + \psi_1^\nu \psi_2^\mu -
(2/3) \, (\psi_1 \cdot \psi_2) \, (P^\mu P^\nu/P^2- g^{\mu \nu})$. Finally,
$\chi_M$ and $\chi_2$ stand for the ratia
\beq
\chi_M ={\alpha_M \over \alpha_0}~,~~~~~ \chi_2 ={\alpha_2 \over \alpha_0}
\label{chis}
\eeq
and encode the relative magnitude of the $O(v^2/c^2)$ relativistic effects due
to respectively the chromo-magnetic interaction (Eq.(\ref{m1})) and the $^3D_1 -
^3S_1$ mixing.

The three $D$ waves correspond to different angular correlations. The first one,
$D_1$, given by Eq.(\ref{d1}) corresponds to a $D$-wave motion in the c.m. frame
of two pions, which correlates with the motion of the c.m. system in the
laboratory frame, i.e. with the direction of ${\vec q}$. This $D$ wave arises in
the leading nonrelativistic approximation~\cite{ns} and is in fact observed and
measured experimentally~\cite{bes} for the transition $\pp \to \pi^+ \pi^- \jp$.
The second $D$-wave amplitude, $D_2$ in Eq.(\ref{d2}), corresponds to the two
pions being in the $S$ wave in their c.m. system and describes the correlation
of the spins of the initial and the final resonances with the $D$-wave motion of
the two-pion system as a whole. Finally, the amplitude $D_3$ given by
Eq.(\ref{d3}) corresponds to a $D$-wave motion of the pions in their c.m. frame,
which $D$ wave is correlated with the spins of the quarkonium states.
It can be noted that the two latter amplitudes are proportional to a product of
two relatively small parameters $\kappa$ and $\alpha_2 + (3/2)\, \alpha_M \sim
v^2/c^2$. Neither $D_2$ nor $D_3$ have yet been observed experimentally,
although a study~\cite{argus} of polarization effects in the decay $\ypp \to
\pi^+ \pi^- \yp$, utilizing a transversal polarization of the DORIS beams
qualitatively confirms that these spin-dependent amplitudes are quite small. (A
discussion can be found in the review \cite{vzai}.)

The transitions between $^1S_0$ states of quarkonium have not been observed yet.
One may hope however that with a dedicated effort a two-pion transition from the
recently found $\eta_c(2S)$ resonance: $\eta_c(2S) \to \pi^+ \pi^- \eta_c$ can
be observed and studied. Within the approach discussed here such transition is
closely related to the familiar decay $\pp \to \pi^+ \pi^- \jp$, and in fact can
be used for a useful calibration of the total width of $\eta_c(2S)$~\cite{mv4}.
Clearly, on the theoretical side the transitions between $^1S_0$ states are
simpler than those between the $^3S_1$ ones since no polarization effects are
involved. On the other hand the effect of the M1 interaction (Eq.(\ref{m1})) is
enhanced for the $^1S_0$ states (Eq.(\ref{am1})) by a factor of 3, so that the
relevant transition amplitudes of a generic $\eta_2 \to \pi^+ \pi^- \eta_1$
transition are given by
\begin{eqnarray}
\label{sep}
&&S(\eta_2 \to \pi^+ \pi^- \eta_1) = \\ \nonumber
&&{4 \pi^2 \over b} \, \alpha_0^{(12)} \, \left [ (1-3 \, \chi_M) \, (q^2+m^2)-
(1+ 3 \, \chi_M) \, \kappa \, \left (1+{2 m^2 \over q^2} \right)\,  \left (
{(q\cdot P)^2 \over P^2} - {1 \over 2} \, q^2 \right ) \right ]~,
\end{eqnarray}
and
\beq
D_1(\eta_2 \to \pi^+ \pi^- \eta_1)= {4 \pi^2 \over b} \, \alpha_0^{(12)} \, (1+3
\, \chi_M) \, {3 \kappa \over 2} \, {\ell_{\mu \nu} P^\mu P^\nu \over P^2}~.
\label{dep}
\eeq
The implications of the enhancement of the relativistic term for the spectrum of
the dipion invariant masses is discussed in the next section.

\section{Phenomenological analysis}
The most experimentally studied so far transitions of the discussed type are
$\pp \to \pi^+ \pi^- \jp$ and  $\ypp \to \pi^+ \pi^- \yp$. In particular, in the
BES experiment~\cite{bes} both the spectrum of the invariant masses of the
two-pion system and the angular distribution described by the $D$ wave amplitude
of the type $D_1$ (Eq.(\ref{d1}) were analyzed, using the formulas equivalent to
Eq.(\ref{spp}) and Eq.(\ref{d1}) with $\chi_M =0$, and the parameter $\kappa$
was determined from the fits. The fit to the mass spectrum resulted in the value
$\kappa=0.186 \pm 0.003 \pm 0.006$, while the fit to the ratio of the $D_1$ and
$S$ wave amplitude gave $\kappa=0.210 \pm 0.027 \pm 0.042$. Clearly, the
consistency of these two values implies that the discussed approach correctly
predicts~\cite{ns} the ratio of the $D_1$ wave in terms of the sub-dominant term
proportional to $\kappa$ in the $S$ wave amplitude. Furthermore, considering the
modification of the expressions for these amplitudes due to the relativistic
parameter $\chi_M$, it is clear that the fit parameter in the experimental
analysis is not exactly $\kappa$ but rather
\beq
\kappa_{eff}^{(\psi' \jp)}={1+\chi_M^{(\psi' \jp)} \over 1-\chi_M^{(\psi' \jp)}}
\, \kappa \approx (1+2\, \chi_M^{(\psi' \jp)}) \, \kappa~.
\label{kappaeff}
\eeq
In the estimates in this section we use the value with the smaller error:
$\kappa_{eff}^{(\psi' \jp)} = 0.186 \pm 0.003 \pm 0.006$.

A similar analysis~\cite{cleo1} of the decay $\ypp \to \pi^+ \pi^- \yp$ was
inconclusive regarding the $D_1$ wave amplitude, whereas the fit for the dipion
invariant mass distribution resulted in $\kappa_{eff}^{(\Upsilon'
\Upsilon)}=0.146 \pm 0.006$. Thus the data display a statistically significant
decrease of the parameter $\kappa_{eff}$ in bottomonium in comparison with the
similar parameter in charmonium: $\kappa_{eff}^{(\psi' \jp)} /
\kappa_{eff}^{(\Upsilon' \Upsilon)} -1 =0.27 \pm 0.07$. Some decrease of this
type of the parameter $\kappa$ was in fact predicted~\cite{ns} based on the
different characteristic scale for normalization of the gluonic operator in
Eq.(\ref{kapf}) relevant for the transitions in the two systems. Indeed, the
characteristic size of $\Upsilon$, $r_\yp \sim 1\,{\rm GeV}^{-1}$, is somewhat
smaller than that of $\jp$, $r_\jp \sim (0.7\, {\rm GeV})^{-1}$. However this
effect is rather difficult to quantify, given that at both these scales the
applicability of the perturbation theory in QCD is marginal. Considering also
that the difference in the size is not large enough, especially in the
logarithmic scale, in order to explain the observed difference in the values of
$\kappa_{eff}$, one may explore another approach to the observed difference by
neglecting the variation of the parameter $\kappa$ altogether and ascribing the
observed effect to different values of the relativistic term $\chi_M$ in these
two quarkonium systems.

Indeed, the relativistic parameter $v^2/c^2$ in the lowest states of bottomonium
is only about one third of that in charmonium, as can be inferred e.g. by
comparing the excitation energies relative to the mass:
$(M_{\pp}-M_{\jp})/(M_{\pp}+M_{\jp}) \approx 3 \,
(M_{\ypp}-M_{\yp})/(M_{\ypp}+M_{\yp})$. Thus one can also expect $\chi_M^{(\psi'
\jp)} \approx 3 \, \chi_M^{(\Upsilon' \Upsilon)}$, and assuming that the value
of $\kappa$ in Eq.(\ref{kapf}) is the same for the transitions in both
quarkonia, one can estimate the relativistic effect of the M1 interaction from
\beq
{\kappa_{eff}^{(\psi' \jp)} \over \kappa_{eff}^{(\Upsilon' \Upsilon)}  } -1
\approx 2\, \chi_M^{(\psi' \jp)} - 2 \chi_M^{(\Upsilon' \Upsilon)} \approx 0.27
\pm 0.07~.
\label{kest}
\eeq
Taken at face value, such estimate gives $\chi_M^{(\psi' \jp)} \approx 0.20 \pm
0.05$ if $\chi_M^{(\Upsilon' \Upsilon)}$ is set equal to ${1 \over 3} \,
\chi_M^{(\psi' \jp)}$ and in $\chi_M^{(\psi' \jp)} \approx 0.14$ if
$\chi_M^{(\Upsilon' \Upsilon)}$ is neglected altogether. In either case the
difference between the specific numerical estimates is within the experimental
and theoretical uncertainty, while the estimated value compares well with the
expected magnitude of the relativistic effects in charmonium.

\begin{figure}[ht]
  \begin{center}
    \leavevmode
    \epsfxsize=10cm
    \epsfbox{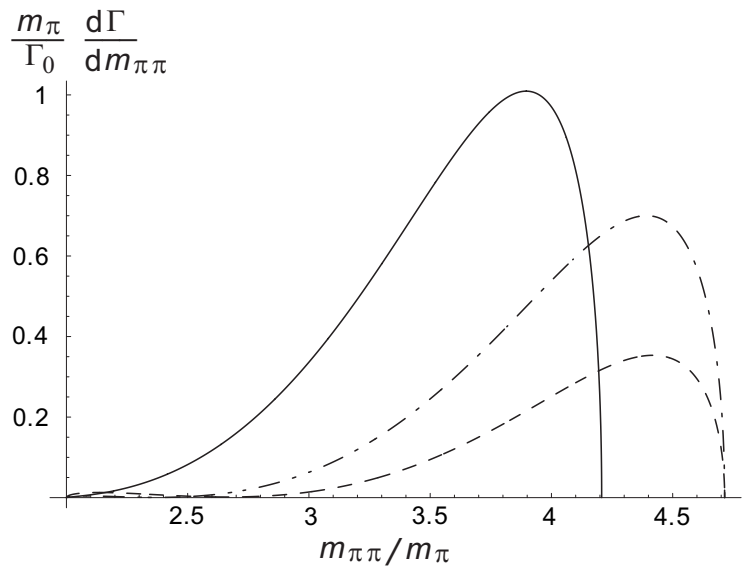}
    \caption{The spectrum of the invariant masses of the two-pion system
described by the equations (\ref{spp}) and (\ref{d1}) for the decay $\pp \to \pi
\pi \jp$ (solid line) and for the decay $\eta_c(2S) \to \pi \pi \eta_c$ with
$\chi_M=0.15$ (dashed) and $\chi_M=0.2$ (dash-dot). The rates are normalized to
the total rate $\Gamma_0$ of the decay $\pp \to \pi \pi \jp$.}
  \end{center}
\end{figure}

A more definite determination of the parameter $\chi_M^{(\psi' \jp)}$ could be
enabled by observing the distribution in the dipion mass in the decay
$\eta_c(2S) \to \pi^+ \pi^- \eta_c$. Indeed, as previously mentioned, the effect
of the chromo-magnetic M1 interaction is enhanced in this decay by a factor of
3. If the value of $\chi_M$ for the charmonium transitions is in the estimated
range $0.15 - 0.2$, the effective parameter $\kappa_{eff}^{(\eta_c' \eta_c)}$
for the latter decay can amount to 0.3 - 0.45. Although the prospect of
measuring the $D_1$ wave in the decay is likely very remote, the effect of such
larger value of $\kappa_{eff}$ should be visible in the more experimentally
accessible spectrum of the dipion mass, as illustrated in Fig.1. As can be seen
from the plots the larger value of $\kappa_{eff}^{(\eta_c' \eta_c)}$ results in
a significant suppression of the spectrum at low invariant mass, due to the zero
of the $S$ wave amplitude at a higher, than for $\pp \to \pi^+ \pi^- \jp$, value
of $q^2$. Also the total rate contains an overall suppression due to the factor
$(1-3 \, \chi_M)^2$ vs. the factor $(1-\chi_M)^2$ for the transition between the
$^3S_1$ states.

\section{Effects of the final state interaction between pions}

So far the amplitudes of the two-pion transitions were considered here in the
chiral limit. The formulas in Eq.(\ref{kapf}) and in Eqs.(\ref{sw}) and
(\ref{dw}) are exact in the leading chiral order, i.e. as far as the quadratic
terms in the pion momenta and mass are concerned. The only dynamical
modification of these expressions can arise from the previously mentioned QCD
renormalization effects. In particular these expressions get no corrections due
to the final state interaction (FSI) between the pions. The latter interaction
however can give rise to the terms whose expansion starts with the fourth power
of momenta and the pion mass, and generally can modify the amplitude at momenta
of the pions relevant for actual transitions in quarkonium. The effects of FSI
in chiral treatment of the two-pion transitions in quarkonium were a matter of
concern ever since the earlier theoretical analyses~\cite{bc}. The effect in the
phases of the amplitudes is well known: these phases for the production
amplitudes are equal to the two-pion scattering phases in the corresponding
partial waves: $S=|S| \, \exp(i \delta_0)$, $D=|D| \, \exp(i \delta_2)$, where
the $I=0$ phases for the $S$ wave, $\delta_0$, and for the $D$ wave, $\delta_2$
are quite well studied.\footnote{It can be mentioned that the
analysis~\cite{bes}
of the data on the $\jp \to \pi^+ \pi^- \jp$ decay does not take into account
the relative phase between the $S$ and $D$ wave pion production amplitudes. Thus
it would be interesting to know whether including the phase factor in the
angular analysis, produces a significant impact on the results.} It is also
generally estimated both on theoretical and phenomenological grounds that the
FSI corrections are not big (at most 20 - 25\%) in the transitions $\pp \to \pi
\pi \jp$ and $\ypp \to \pi \pi \yp$. (For a discussion see the
review~\cite{vzai}.) Some phenomenological arguments in favor of such estimate
will also be discussed further towards the end of this section, and we start
with  a theoretical estimate of the onset of the higher term in the chiral
expansion.

The interaction of pions at low energy in the $D$ wave is quite weak, so that
any modification by FSI of the $D$ wave production amplitude of Eq.(\ref{dw})
can be safely neglected, and only the modification of the $S$ wave amplitude
(\ref{sw}), $\delta S_{\mu \nu \lambda \sigma}$, is of interest for present
phenomenology. The imaginary part of the correction at $q^2 > 4m^2$ is found
from the unitarity relation in terms of the isospin I=0 $\pi \pi \to \pi \pi$
scattering amplitude $T(q^2)$ in the $S$ wave as
\beq
{\rm Im}\, ( \delta S_{\mu \nu \lambda \sigma}) = \sqrt{1-{4m^2 \over q^2}}\,
{T(q^2) \over 16 \pi} \, S_{\mu \nu \lambda \sigma} ~.
\label{unit}
\eeq
The amplitude $T(q^2)$ is well known in the chiral limit, i.e. in the quadratic
in $q$ and $m$ approximation, since the work of Weinberg~\cite{weinberg}. In the
normalization used here this amplitude has the form
\beq
T(q^2)={2 \, q^2 - m^2 \over f_\pi^2}~,
\label{tpipi}
\eeq
where $f_\pi \approx 130\,$MeV is the $\pi^+ \to \mu^+ \nu$ decay constant.
Clearly, the expression in Eq.(\ref{unit}) is of the fourth power in $q$ and
$m$.

The real part of $\delta S_{\mu \nu \lambda \sigma}$ can then be estimated  from
Eq.(\ref{unit}) using the dispersion relation in $q^2$ for the amplitude $S$. In
doing so one should set the condition for the subtraction constants that this
real part does not contain quadratic (and certainly also constant) terms in $q$
and $m$, since these are given by Eq.(\ref{sw}). After these subtractions the
remaining dispersion integral is still logarithmically divergent and contains
the well known `chiral logarithm', depending on the ultraviolet cutoff
$\Lambda$, which is usually set at $\Lambda \sim 1\,$GeV, i.e. the scale where
any chiral expansion certainly breaks down. Using the equations (\ref{sw}),
(\ref{unit}) and (\ref{tpipi}) one can readily estimate the first FSI correction
with a `logarithmic accuracy'. The expression for the full $S$ wave production
amplitude then takes the form
\begin{eqnarray}
\label{sw2}
&&S_{\mu \nu \lambda \sigma}  =  {8 \pi^2 \over 3 b} \, \left \{ (q^2+m^2)\,
(1+\xi_1)\, (g_{\mu \lambda} g_{\nu \sigma} - g_{\mu \sigma} g_{\nu \lambda})
-\right .
\\  \nonumber
&&\left .  {3 \over 4} \, \kappa \, \left ( 1+ {2 m^2 \over q^2} \right ) \, (1
+  \xi_2) \,\left [ q_\mu q_\lambda g_{\nu \sigma}+q_\nu q_\sigma g_{\mu
\lambda}-q_\nu q_\lambda g_{\mu \sigma}-
q_\mu q_\sigma g_{\nu \lambda}-{1 \over 2} \, q^2\,(g_{\mu \lambda} g_{\nu
\sigma} - g_{\mu \sigma} g_{\nu \lambda})\right ] \right \}~,
\end{eqnarray}
where the correction terms $\delta_1$ and $\delta_2$ to Eq.(\ref{sw}) are given
as
\beq
\xi_1 = {2 \, (q^2)^2- 7 \, q^2 \, m^2 +3 \, m^4 \over 16 \pi^2 \, f_\pi^2 \,
(q^2+m^2)}\, \ln {\Lambda^2 \over m^2}+ {\rm i} \, {2 \, q^2 - m^2 \over 16 \pi
\, f_\pi^2}\, \sqrt{1-{4m^2 \over q^2}} ~,
\label{xi1}
\eeq
and
\beq
\xi_2 = {2 \, (q^2)^2- 9 \, q^2 \, m^2 +8 \, m^4 \over 16 \pi^2 \, f_\pi^2 \,
(q^2+2 m^2)}\, \ln {\Lambda^2 \over m^2}+{\rm i} \, {2 \, q^2 - m^2 \over 16 \pi
\, f_\pi^2} \, \sqrt{1-{4m^2 \over q^2}}~,
\label{xi2}
\eeq
where the non-logarithmic imaginary part is retained for reference regarding the
normalization. The lower limit under the logarithm is generally a function of
both $q^2$ and $m^2$, however any difference of this function from the value
$m^2$ used in Eqs.(\ref{xi1}) and (\ref{xi2}) is a non-logarithmic term, i.e.
beyond the accuracy of these equations. Since $m^2$ is the smallest of the two
parameters in the physical region of pion production, it can be expected that
using this parameter provides a conservative estimate of the effect of FSI.

Estimating the corrections in Eq.(\ref{xi1}) and Eq.(\ref{xi2}), one finds that
at the lower end of the physical phase space, i.e. near $q^2=4 \, m^2$, these
terms do not exceed few percent. Thus the corrections only weakly modify the
normalization of the pion production amplitude near the threshold. A theoretical
extrapolation to higher values of $q^2$ is problematic, and, most likely, one
would have to resort to using actual data on the dipion spectra in order to
judge upon the significance of deviation from the essentially linear in $q^2$
behavior of the amplitude described by Eq.(\ref{spp}). A quantitative estimate
of the deviation from this behavior has been attempted~\cite{vzai} using the
data~\cite{argus} on $\ypp \to \pi^+\pi^- \yp$ by parametrizing the deviation as
a factor $(1+q^2/M^2)$ in the amplitude with $M$ being a parameter. The thus
obtained lower limit on $M$ is $1\,$GeV at 90\% C.L.

Another phenomenological argument in favor of a relatively moderate FSI effect
in the absolute value of the dominant $S$ wave in the decay $\pp \to \pi^+ \pi^-
\jp$ stems from the previously mentioned agreement of the observed~\cite{bes}
value of the ratio $D/S$ with the parameter $\kappa$ entering the expression for
the $S$ wave and extracted from the two-pion invariant mass spectrum. Clearly
such an agreement would be ruined if there was a significant enhancement of the
$S$ wave by FSI.

Furthermore, the observed ratio of the rates of the transitions  $\pp \to \eta
\jp$ and $\pp \to \pi^+ \pi^- \jp$ reasonably agrees with the
calculation~\cite{vz}, which neglects any FSI effects in the latter decay, while
it is clear that there is no such effects in the emission of one-particle, i.e.
of $\eta$. The expected accuracy of the theoretical calculation is mostly that
of applying the nonrelativistic limit to charmonium, i.e. $O(20\%)$ in the
amplitude.
Thus the theoretical (and experimental) uncertainty may allow for a presence of
the FSI effects in the two-pion system at such level, however large effects of
this type, like those recently claimed in Ref.~\cite{guo}, are definitely
excluded.

Finally, one can notice in connection with the equations (\ref{d2}) and
(\ref{d3}) for the $D_2$ and $D_3$ waves involving the polarizations of the
quarkonium resonances, that the unknown quarkonium matrix elements all cancel in
the ratio $D_2/D_3$. The $D_2$ amplitude is determined by the $S$ wave of the
pions in their c.m. system, and the $D_3$ contains the pions in the $D$ wave
relative to each other. Thus if the ratio of actual amplitudes $D_2/D_3$ could
be measured from angular distributions, e.g. in the decay $\pp \to \pi^+ \pi^-
\jp$, this would produce direct data on the modification by FSI of the $S$ wave
relative to the $D$ wave.

\section{Resolving the $\yppp \to \pi \pi \yp$ puzzle?}
The decay $\yppp \to \pi \pi \yp$ is known to be quite different from the ``well
behaved" transitions $\pp \to \pi^+ \pi^- \jp$ and $\ypp \to \pi^+ \pi^- \yp$ in
that the spectrum~\cite{cleo2} of the dipion invariant masses in this decay has
two distinct maxima at low and high values of $q^2$. The proposed solutions to
the puzzle presented by this behavior included presence of  a hypothetical
exotic resonance~\cite{mv6,bdm}, breakdown of the multipole
expansion~\cite{moxhay}, and unusual FSI effects~\cite{ckk}. In this section I
exercise an alternative, and somewhat more conventional explanation of the
observed behavior of the spectrum in terms of the considered here terms of the
multipole expansion. Namely, one can readily verify that the dipion mass
spectrum in the decay $\yppp \to \pi \pi \yp$ is reasonably reproduced by the
equations (\ref{spp}) - (\ref{d3}) if the parameters $\chi_M$ and $\chi_2$ for
this decay are rather large: of order one, which is probably a result of a
dynamical suppression of the leading nonrelativistic quarkonium amplitude
$\alpha_0$.

The overall amplitude of the decay $\yppp \to \pi \pi \yp$ is arguably strongly
suppressed. Indeed, the observed absolute rate of the transition  $\yppp \to
\pi^+ \pi^- \yp$ is only about 0.2 of the rate of $\ypp \to \pi^+ \pi^- \yp$ in
spite of a significantly larger available phase space. The estimated suppression
of the former decay would be even stronger, if its amplitude was "well behaved",
i.e. given by leading E1 interaction and thus dominantly proportional to $q^2$.
Furthermore, the total rate of another similar transition from the same $\yppp$
resonance, $\yppp \to \pi \pi \ypp$, is about 0.6 of the rate of $\yppp \to \pi
\pi \yp$, even though the energy released in the former transition is only
slightly above the two-pion threshold, and the transition is very strongly
kinematically suppressed. Thus it appears quite reasonable to conclude that the
pure $S$ wave part of the quarkonium amplitude in Eq.(\ref{aij}), $\alpha_0$ is
small for the discussed transition. This can be a result of cancellation in the
(appropriately weighed) overlap of the wave functions determining this amplitude
due to the oscillations of the $3S$ wave function. Such cancellation however is
not necessarily present in the overlap terms due to the $^3D_1-^3S_1$ mixing,
$\alpha_2$ in Eq.(\ref{asd}), and due to the M1 interaction, $\alpha_M$ in
Eq.(\ref{amg}). The latter terms are naturally expected to be of the order
$O(v^2/c^2)$ in comparison with an unsuppressed leading amplitude, e.g. the
amplitudes of the ``well behaved" transitions: $\ypp \to \pi \pi \yp$ or $\yppp
\to \pi \pi \ypp$. In other words, the ratia $\chi_M$ and $\chi_2$
(Eq.(\ref{chis})) can be large due to the small denominator $\alpha_0$.
A fully quantitative analysis of the decay $\yppp \to \pi \pi \yp$ is
complicated by that the energy released in the transition $W=M_{\yppp}-M_{\yp}
\approx 895\,$MeV is rather large for a straightforward application of the
chiral-limit formulas for the pions. However, for at least a qualitative
estimate, I neglect here this complication and apply the equations (\ref{spp}) -
(\ref{d3}) to evaluate the dipion mass spectrum. The resulting behavior is
illustrated in Fig.2, where the parameters are set as $\chi_M=0.7$,
$\chi_2=1.0$, and $\kappa=0.13$. Clearly, the evaluated dipion mass spectrum
closely resembles the experimentally observed~\cite{cleo2}, although no attempt
is made here at a quantitative fit to experimental data.

\begin{figure}[ht]
  \begin{center}
    \leavevmode
    \epsfxsize=10cm
    \epsfbox{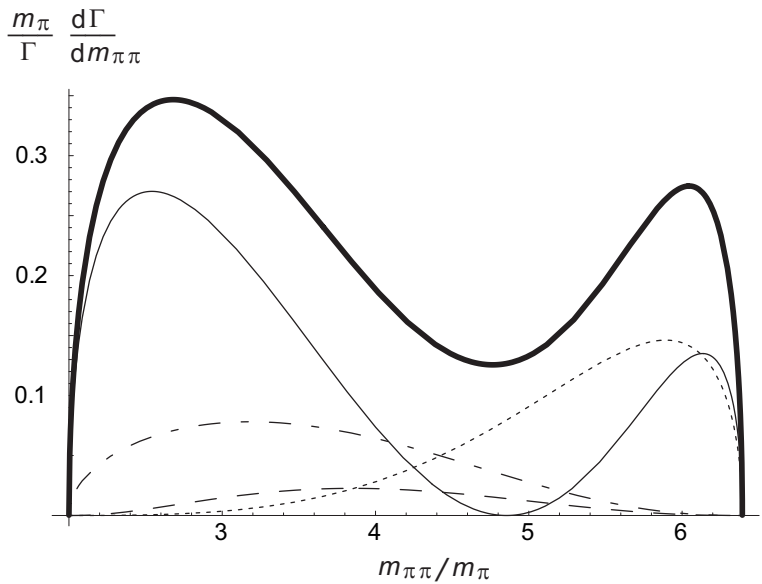}
    \caption{The spectrum of the invariant masses of the two-pion system as
given by the equations (\ref{spp}) - (\ref{d3}) for the decay $\yppp \to \pi \pi
\yp$. Shown in the plot are the total distribution (thick line), and the
contribution of individual partial waves:  $S$ (thin solid line), $D_1$
(dashed), $D_2$ (dash-dot), and $D_3$ (dotted). }
  \end{center}
\end{figure}

In order to assess whether the values of the relativistic parameters $\chi_M$
and $\chi_2$ used in the plots of Fig.2 are of the order of the expected
relativistic effects in bottomonium, it is instructive to compare the
corresponding amplitudes $\alpha_M^{(\yp'' \yp)}$ and $\alpha_2^{(\yp'' \yp)}$
with the amplitude of the transition, that appears to be `normal', namely $\yppp
\to \pi \pi \ypp$. In the latter transition the released energy is only
332\,MeV, and according to the discussed approach it should be absolutely
dominated by the $S$ wave amplitude given by Eq.(\ref{spp}). Also any FSI
effects in this decay should be very small due to the proximity of the pions to
the threshold. To certain extent this approach can be tested by comparing the
rates of the transitions with the charged pions and with the neutral, by using
respectively the mass of the charged and neutral pion in considering each of
these decay modes. The effect of the pion mass difference is quite essential due
to the small available energy, so that after numerical integration of the rate
calculated from Eq.(\ref{spp}) one finds, in place of the isotopic ratio 2, the
estimate
\beq
{\Gamma[\yppp \to \pi^+ \pi^- \ypp] \over \Gamma[\yppp \to \pi^0 \pi^0 \ypp]}
\approx 1.26~,
\label{yppprat}
\eeq
which is in the agreement with the data~\cite{pdg}: ${\cal B}[\yppp \to \pi^+
\pi^- \ypp] = (2.8 \pm 0.6)\%$ and ${\cal B}[\yppp \to \pi^0 \pi^0 \ypp] = (2.00
\pm 0.32)\%$.~\footnote{In this estimate the value of $\kappa_{eff}$ used is
0.146. Due to the small energy in the process, the estimate is quite insensitive
to this particular value as long as $\kappa_{eff}$ is small.}
The integral over the spectrum of either of the modes of transition $\yppp \to
\pi \pi \ypp$ can then be compared with the result of the numerical integration
over the spectrum of the decay $\yppp \to \pi^+ \pi^- \yp$ produced by the
amplitudes in the equations (\ref{spp}) - (\ref{d3}) which gives the ratio of
the rates of the two transitions from $\yppp$ in terms of the ratio of the
squares of the corresponding amplitudes $\alpha_0$ and can be compared with the
data. Performing this calculation with the values of the parameters
$\chi_M^{(\yp'' \yp)}=0.7$ and $\chi_2^{(\yp'' \yp)}=1.0$ used in the plots of
Fig.2, gives the estimate
\beq
{\alpha_0^{(\yp'' \yp)} \over \alpha_0^{(\yp'' \yp')}} \approx {\alpha_2^{(\yp''
\yp)} \over \alpha_0^{(\yp'' \yp')}} \approx 0.06~,~~~~~~ {\alpha_M^{(\yp''
\yp)} \over \alpha_0^{(\yp'' \yp')}} \approx 0.04~,
\label{aest}
\eeq
which certainly falls in the range of naturally expected magnitude of the
relativistic terms in bottomonium: $v^2/c^2 \sim (M_{\ypp}- M_{\yp})/(M_{\ypp}+
M_{\yp}) \approx 0.06$, and also quantifies the dynamical suppression of the
leading nonrelativistic amplitude $\alpha_0$ in the transition $\yppp \to \pi
\pi \yp$. However, given the large energy release in this transition it is
rather difficult to quantify at present the FSI effect on the estimate
(\ref{aest}).  In either event, the suggested explanation of the observed
peculiarity of the decay $\yppp \to \pi \pi \yp$, heavily relies on the relative
prominence of the spin-dependent $D$ waves, $D_2$ and $D_3$, in this decay, as
can be seen from the plots of Fig.2. The relative contribution of these two
waves is the largest near the minimum in the invariant mass distribution at
$m_{\pi \pi} \approx 0.65\,$GeV, which is the main experimentally testable
qualitative prediction of the suggested mechanism.

It should be mentioned in connection with the suggested presence of the
spin-dependent $D$ waves that an experimental search for quarkonium polarization
effects in $\yppp \to \pi \pi \yp$ has been done~\cite{cleo2}. It was found that
the data were consistent at 65\% confidence level ``with the expectation that
the daughter $\Upsilon(1S)$ retains the polarization of the parent $\yppp$ along
the beam axis". It thus appears that the test has not been sensitive enough and
a more detailed study of the bottomonium polarization effects in this transition
is needed.

\section{Summary}
In summary. Within the standard approach to two-pion transitions in heavy
quarkonium, based on the multipole expansion in QCD and chiral dynamics of the
pions, the first relativistic terms of order $v^2/c^2$ in the transition
amplitudes arise from the effects of the $^3D_1  -^3S_1$ mixing in the part
determined by the E1 interaction and from the chromo-magnetic M1 interaction.
These terms are parametrized by the quantities $\alpha_2$ (Eq.(\ref{asd})) and
$\alpha_M$ (Eq.(\ref{amg})). The significance of the chromo-magnetic term in
two-pions transitions in charmonium can be approximately estimated from the
available data on the dipion invariant mass spectrum in the decays $\pp \to \pi
\pi \jp$ and $\ypp \to \pi \pi \yp$. The estimated value is 0.15 - 0.2 of the
dominant nonrelativistic amplitude. The effect of the chromo-magnetic term is
enhanced in the yet unobserved transition $\eta_c(2S) \to \pi \pi \eta_c$ and is
expected to significantly distort the spectrum in the latter decay and also
result in a suppression of its rate, as shown in Fig.1. The absolute
determination of the leading nonrelativistic amplitude in charmonium, the
chromo-polarizability, is of interest for other applications, e.g. the
charmonium scattering on nuclei. Such determination generally suffers from FSI
effects of the two pion rescattering. These effects are estimated as a next term
in the chiral expansion and amount to only few percent at a low invariant mass
of the two-pion system near the threshold. An extrapolation to higher values of
$q^2$ can be done using experimental data. With the presently available data
there is no indication of  a large FSI effect. Furthermore, an agreement of the
chiral-limit formulas with the data~\cite{bes} on the $D$ wave in the decay $\pp
\to \pi \pi \jp$, as well as the agreement with the data of the theoretical
prediction~\cite{vz} for the ratio of the rates of the decays $\pp \to \eta \jp$
and $\pp \to \pi \pi \jp$, suggest that the FSI effects may amount to at most a
moderate fraction of the amplitude of the transition in charmonium. An
experimental measurement of the ratio of the $D_2$ wave (Eq.(\ref{d2})) and the
$D_3$ wave (Eq.(\ref{d3})) in $\pp \to \pi \pi \jp$ would provide a direct test
of the FSI effect. The relativistic terms in the two-pion transitions may hold
the clue to solving the puzzle of the unusual dipion mass spectrum observed in
the transition $\yppp \to \pi \pi \yp$, if the leading nonrelativistic
quarkonium matrix element in this transition is strongly suppressed due to
details of bottomonium wave functions. Although a detailed quantitative
description of the latter decay is not yet attainable within the present
knowledge, the suggested mechanism necessarily predicts a noticeable presence of
the polarization-dependent $D_2$ and $D_3$ waves in the amplitude, which
prediction can be tested experimentally.

\section*{Acknowledgements}
I thank Misha Shifman and Arkady Vainshtein for illuminating discussions.
This work is supported in part by the DOE grant DE-FG02-94ER40823.

\end{document}